\documentclass[usenatbib]{mn2e}

\usepackage{graphicx}
\usepackage[space]{grffile}
\usepackage{latexsym}
\usepackage{amsfonts,amsmath,amssymb}
\usepackage{url}
\usepackage[utf8]{inputenc}
\usepackage{hyperref}
\hypersetup{colorlinks=false,pdfborder={0 0 0}}
\usepackage{textcomp}
\usepackage{longtable}
\usepackage{multirow,booktabs}
\usepackage{hhline}

\usepackage{natbib}

\title[Non-linear structure formation in the ``Running FLRW'' cosmological model]{Non-linear structure formation in the ``Running FLRW'' cosmological model}

\author[Antonio Bibiano]{Antonio Bibiano$^1$, Darren J. Croton$^1$\\
$^1$ Centre for Astrophysics and Supercomputing, Swinburne University of Technology, Hawthorn, Victoria 3122, Australia}

\begin{document}

\maketitle

\label{firstpage}

\begin{abstract}
We present a suite of cosmological N-body simulations describing the ``Running Friedmann-Lema{\"i}tre-Robertson-Walker'' (R-FLRW) cosmological model. This model is based on quantum field theory in a curved space-time and extends $\Lambda$CDM with a time-evolving vacuum density, ${\Lambda(z)}$, and time-evolving gravitational Newton's coupling, ${G(z)}$. In this paper we review the model and introduce the necessary analytical treatment needed to adapt a reference N-body code.
Our resulting simulations represent the first realisation of the full growth history of structure in the R-FLRW cosmology into the non-linear regime, and our normalisation choice makes them fully consistent with the latest cosmic microwave background data. The post-processing data products also allow, for the first time, an analysis of the properties of the halo and sub-halo populations. We explore the degeneracies of many statistical observables and discuss the steps needed to break them. Furthermore, we provide a quantitative description of the deviations of R-FLRW from $\Lambda$CDM, which could be readily exploited by future cosmological observations to test and further constrain the model.
\\
\\
\\

\end{abstract}


\bibliographystyle{mn2e_eprint}
\providecommand{\eprint}[1]{\href{http://arxiv.org/abs/#1}{arXiv:#1}}
\providecommand{\adsurl}[1]{}

\providecommand{\ISBN}[1]{\href{http://cosmologist.info/ISBN/#1}{ISBN: #1}}

\section{Introduction}

One of the most significant achievements of modern cosmology has been our ability to simultaneously fit a wide range of independent cosmological observations with a simple parametrised model at a remarkable level of precision. Examples of such observations include the recent Planck mission's Cosmic Microwave Background Radiation (CMBR) measurements \citep{Ade:2015xua}, the distance measurements for Type Ia supernovae \citep{panstarr}, and the distribution of galaxies in the large-scale structure of the Universe (see e.g. \citealt{wigglez}, \citealt{bosspaper}). All of these observations can be described by the simple $\Lambda$CDM model, and in turn provide increasingly stringent constraints on the model's parameters, including the so called cosmological constant, $\Lambda$. 

$\Lambda$ was added to the standard equations of general relativity to fit the observations and is treated as an additional component, alongside matter and radiation. It accounts for about 70\% of the energy density in today's Universe. With this simple addition, and with some fine-tuning, we have a powerful model that allows us to make very precise predictions. $\Lambda$CDM gives us a clear picture of (almost) the entire history of the Universe, one that is consistent with being geometrically flat and currently undergoing an epoch of accelerated expansion \citep{carrollCC}. Its main shortcoming is that the model provides little insight into the nature of this additional component, which astronomers and physicists alike find highly unsatisfactory.

In trying to understand this dominant component, the first interpretations speculated that $\Lambda$ might be a manifestation of the energy density of the vacuum. This interpretation, while very natural, is impossible to reconcile with the calculated value of $\Lambda$ arising from any known quantum field theory mechanism \citep{weinbergCC}. Due to this failure $\Lambda$ was dubbed ``dark energy''. Its emergence in cosmology has motivated a whole new area of research aimed at explaining its nature and behaviour.

Many mechanisms have been developed to explain the underlying nature of dark energy while also preserving the successes of the $\Lambda$CDM model. These mechanisms can be roughly divided into two categories. In the first, dark energy can be treated as the manifestation of one or more additional fields. The behaviour of these fields can be described analytically and different models are distinguished by the choice of the potential, as in the various quintessence models \citep{Wetterich88, RatraPeebles} which have become very popular. The fields can also be described in an approximate way by simply adding degrees of freedom to the standard $\Lambda$CDM model, as in the Chevallier--Polarski--Linder \citep{CPLcp, CPLl} and other parametrized models.  We will refer to these kind of models ``dark energy models''.

The second category of models rely on modifying Einstein's equations of general relativity to obtain the accelerated expansion without the need for additional components. These usually can be interpreted as higher-order corrections to general relativity, or completely different gravitational theories. They usually share a common small-scale limit, to agree with all the experiments that confirmed general relativity in our Solar System, while exhibiting a different behaviour on large scales that, in turn, has the same effect of a cosmological constant. An example of such theories are f(R)-gravity \citep{HuSawicki} and Galileon models \citep{galileon}. We refer to these as ``modified gravity models''.

Regardless of its origin, to test such models it is necessary to obtain accurate predictions. While testing against certain observables, like the CMBR and supernovae, a linear treatment is enough, for large-scale structure measurements it is necessary to extend any analysis to the non-linear regime. This regime is usually impossible to probe with analytic theory alone.

The exploration of the non-linear regime in the past decade has predominantly been undertaken with the aid of cosmological N-body simulations. These simulations allow us to study the evolution of the matter distribution in the Universe under the influence of both gravity and cosmic expansion, down to small scales. Their predictive power has led to the development of a whole new branch of cosmology dedicated to the development of algorithms and highly efficient codes necessary to carry out those simulations. Historically, the goal of numerical cosmology has been to reach high precision and large dynamical range for simulations in the $\Lambda$CDM scenario. Only recently has the attention shifted towards different cosmological scenarios.

We have now reached the stage where future surveys, like Euclid \citep{EuclidPap} and eROSITA \citep{eROSITA}, will be able to highlight even small deviations from the $\Lambda$CDM scenario, and will potentially be able to distinguish between different but very close cosmological scenarios. For this reason it is prudent to adapt the standard N-body algorithms and codes to simulate these new scenarios. Many such modifications have already been carried out for various dark energy models, e.g. the works by \cite{baldiCodecs}, \cite{LiQuinte}, and modified gravity models as described in  \cite{ecosmog} and \cite{mggadget}.

In this paper we introduce our own modification to the popular \textsc{gadget} N-body code  \citep{gadget2} to simulate an interesting scenario that falls halfway between the two categories introduced earlier: the ``Running Friedmann-Lema{\"i}tre-Robertson-Walker'' (R-FLRW) cosmological model. This model was introduced by \cite{Grande_2011} and retains the interpretation of the cosmological constant as vacuum energy but allows for the evolution of its energy density. This in turn requires a variation of the gravitational constant with time to retain the local conservation of matter. 

Our paper is organised as follows: In Section \ref{sec:model} we present the theory behind the R-FLRW model. In Section \ref{sec:perturb} we describe matter perturbations in the linear regime under the action of this model. Section \ref{sec:simul} is devoted to a description of our simulation algorithm, while in Section \ref{sec:results} we present the results for a suite of simulations run with this algorithm. We give a discussion and draw our conclusions in Section \ref{sec:conclusions}.

\section{The Model}
\label{sec:model}
The ``Running Friedmann-Lema{\"i}tre-Robertson-Walker'' (R-FLRW) cosmological model was first described in \cite{Grande_2011}. In this model the value of the cosmological constant is seen as an effective quantity whose value evolves with the expansion of the Universe. As a consequence, the model also enforces the conservation of matter by allowing for an evolution in the strength of the gravitational coupling. It is important to highlight that the R-FLRW model retains the standard $\Lambda$CDM model's interpretation of the cosmological constant as vacuum energy while considering the reasonable possibility that its energy density might be related to other time-varying cosmological quantities. This idea has solid roots in fundamental physics, and we refer the reader to the aforementioned literature for a thorough description of the underlying quantum field theory background necessary to justify some the model choices here. In the present work we will limit our discussion to an introduction of the main equations and the notation necessary for our analysis.

To describe the R-FLRW model we start with the standard general relativistic description of the interaction between the curvature of space-time and its matter content, as described by the Einstein equations:
\begin{equation}
    R_{\mu\nu} - \frac{1}{2}g_{\mu\nu}R = 8 \pi G T_{\mu\nu} + \Lambda g_{\mu\nu} ~.
\end{equation}
Here the cosmological constant term $\Lambda$ is interpreted as a source that can be incorporated in the modified energy-momentum tensor by
\begin{equation}
    \tilde{T}_{\mu\nu} \equiv T_{\mu\nu}  +  \frac{\Lambda}{8 \pi G}= (\rho_\Lambda - \rho_m) g_{\mu\nu} + (\rho_m + p_m)U_\mu U_\nu ~.
\end{equation}
This form arises from a description of the matter content of the Universe as a perfect fluid with velocity 4-vector $U_\mu$, and the inclusion of a vacuum energy density $\rho_\Lambda = \frac{\Lambda}{8 \pi G}$ associated with the cosmological constant term. Throughout we also assume a spatially flat Friedmann-Lema{\"i}tre-Robertson-Walker (FLRW) metric,
\begin{equation}
    ds^2 = dt^2 - a^2(t)d\vec{x}^2 ~,
\end{equation}
with scale factor $a(t)$.

The above general framework is also the basis of the standard $\Lambda$CDM cosmological model, but the cosmological principle embodied by the FLRW metric allows $\rho_\Lambda$ and $G$ to be functions of time without losing the covariance of the theory. In fact, the Bianchi identities imply that
\begin{equation}
    \nabla^{\mu}(G \tilde{T}_{\mu\nu}) = 0 ~,
\end{equation}
which in our case becomes
\begin{equation}
	\label{eq:bianchi}
    \frac{d}{dt}[G (\rho_\Lambda + \rho_m)] + 3 G H (\rho_m + p_m) = 0 ~,
\end{equation}
where $H$ is the standard Hubble parameter $H \equiv \dot{a} / {a}$. Equation~\ref{eq:bianchi} implies local conservation of matter when
\begin{equation}
\label{matterconservation}
    \frac{d}{dt}\rho_m + 3 G H (\rho_m + p_m) = 0 ~,
\end{equation}
which will be true in two cases: (1) when $G$ and $\rho_\Lambda$ are constants, as in the standard $\Lambda$CDM model; or (2) when both $G$ and $\rho_\Lambda$ are functions of time and satisfy the constraint
\begin{equation}
\label{bianchiconstraint}
    (\rho_m + \rho_\Lambda) \frac{dG}{dt} + G \frac{d\rho_\Lambda}{dt} = 0 ~.
\end{equation}

From the above, the background expansion of the Universe can be characterised by a key set of functions: $H(t)$, $\rho_m(t)$, $p_m(t)$, $G(t)$ and $\rho_\Lambda(t)$. To obtain a closed system of equations that describes the evolution of these quantities in a R-FLRW Universe we need to specify the functional form for the evolution of $G$ and $\rho_\Lambda$. Following \cite{Grande_2011}, we start by assuming that the gravitational coupling and the vacuum energy density evolves as a power series of some energy scale $\mu$, with rates given by
\begin{align}
\label{variationrates}
    \frac{d\rho_\Lambda(\mu)}{d \text{ ln } \mu} &= \sum\limits_{k =0,1,2,...} A_{2k} \mu^{2k},\\
    \frac{d}{d \text{ ln } \mu}\left( \frac{1}{G(\mu)} \right) &= \sum\limits_{k =0,1,2,...} B_{2k} \mu^{2k} ~.
\end{align}
Although this form is purely phenomenological, a sensible choice for the energy scale is $H$. In this way we associate the running of the cosmological quantities to the typical energy scale of the gravitational field associated with the FLRW metric.

Let us now consider the evolution of $\rho_\Lambda$ using the above. This type of expansion has been widely discussed in the literature, and according to the results of \cite{Basilakos_2009}, \cite{Babi__2002} and \cite{Borges_2008} we only need to keep the zeroth and second orders to prevent deviations from the $\Lambda$CDM model that are too large to be reconciled with current observations. After integration we obtain the functional form
\begin{equation}
\label{firstlambdalaw}
    \rho_\Lambda (H) = n_0 + n_2 H^2 ~,
\end{equation}
with coefficients $n_0$ and $n_2$ given by
\begin{equation}
    n_0 = \rho_\Lambda^0 - \frac{3 \nu}{8 \pi} M_P^2 H_0^2, ~~~ n_2 = \frac{3 \nu}{8 \pi}M_P^2 ~.
\end{equation}
We have simplified the expression above by writing
\begin{equation}
    \nu = \frac{1}{6 \pi } \sum_i B_i \frac{M_i^2}{M_P^2} ~.
\end{equation}
In the usual way, $H_0$ and $\rho_\Lambda$ are the present day values of the Hubble parameter and vacuum energy density, respectively, while $M_i$ is the mass associated with the $i$th term of the expansion in the underlying quantum field theory derivation of the model (Equation~\ref{variationrates}).

It is worth emphasising that the parameter $\nu$ is a critical component of the new framework; when $\nu = 0$ the vacuum energy remains constant with $\rho_\Lambda = \rho_\Lambda^0$, but when $\nu \not= 0$ the evolution law (i.e. Equation~\ref{firstlambdalaw}) can be rewritten as
\begin{equation}
\label{secondlambdalaw}
    \rho_\Lambda(H) = \rho_\Lambda^0 + \frac{3 \nu}{8 \pi} M_P^2 (H^2 - H_0^2) ~.
\end{equation}
In \cite{Grande_2011} $\nu$ was considered a free parameter with a natural range of $|\nu| \ll 1$. More specifically, the range of $\nu$ was constrained against joint supernovae, CMBR and BAO observations to lie in the range $-0.004 < \nu < 0.002$. This ensures the R-FLRW model is consistent with current measurements at at least the 1$\sigma$ level.

\begin{figure}
\includegraphics[width=\columnwidth]{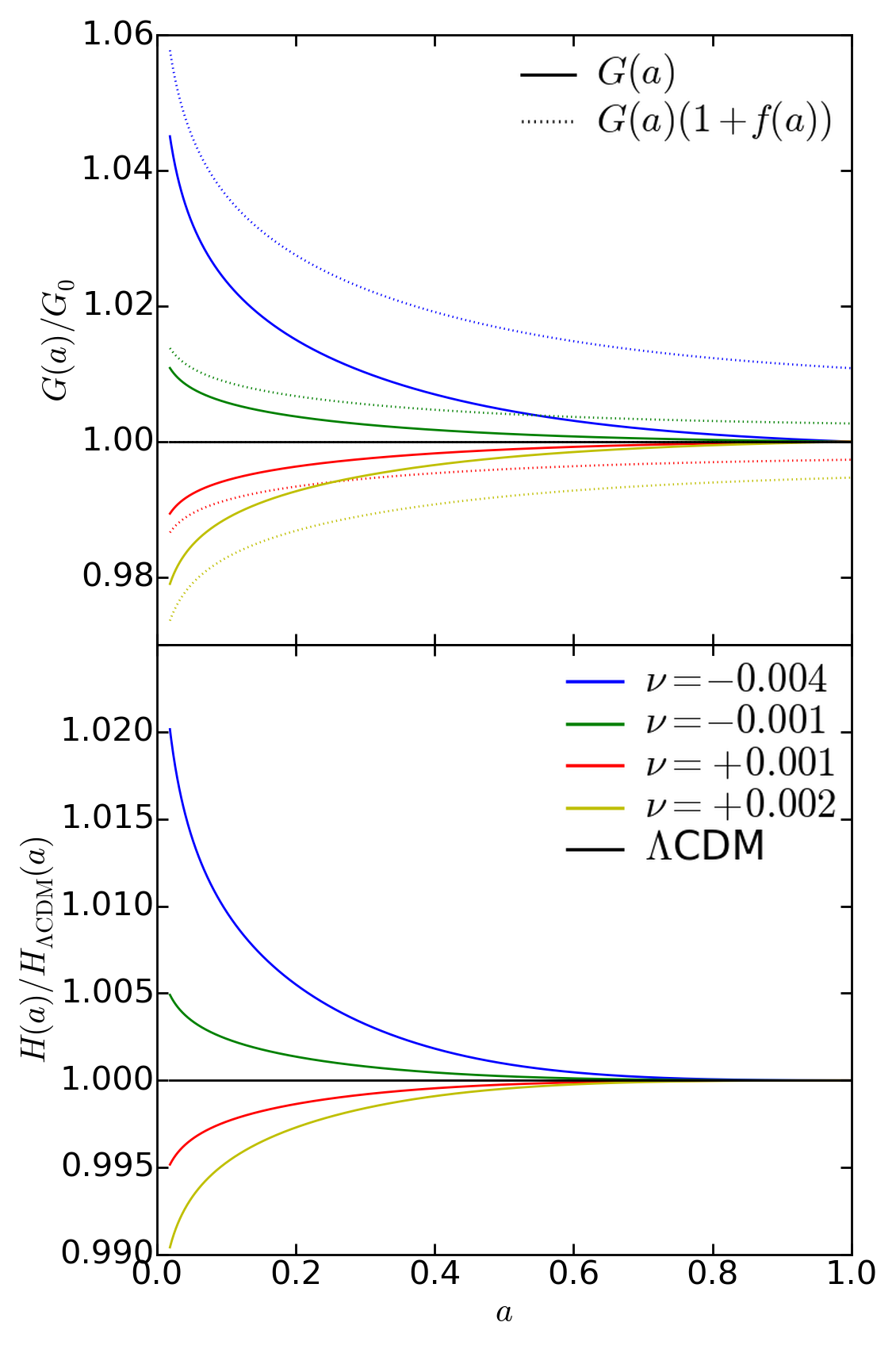}
\caption{A comparison of $G$ and $H$ for different values of the $\nu$ parameter that correspond to the simulated R-FLRW scenarios: in the top panel the solid lines show the time dependence of G(a), while the dotted lines show the time dependence of $\tilde{G}$(a) from Equation \ref{tildegdef}. This includes our proposed approximation and is the value used in the simulation code to include the effects of perturbations in $G$ and $\Lambda$. This correctly reproduces the linear growth of the R-FLRW model to within 0.5\%, as explained in Section \ref{sec:perturb} and shown in the bottom panel of Figure~\ref{fig:growth}.  The bottom panel shows the time dependence of the R-FLRW Hubble function through its ratio with the $\Lambda$CDM Hubble function.}

\label{fig:handg}
\end{figure}

An evolution equation for $G $ can now be obtained by again keeping only the dominant terms in Equation~\ref{variationrates}, and by combining Equation~\ref{secondlambdalaw} for $\rho_\Lambda$ with the constraint given by Equation~\ref{bianchiconstraint} imposed by the Bianchi identity. After integration the solution reads
\begin{equation}
\label{variationofg}
    g(H) \equiv \frac{G(H)}{G_0} = \frac{1}{1+\nu \text{ ln } (H^2/H_0^2)} ~.
\end{equation}
 In the top panel of  Figure \ref{fig:handg} we plot Equation \ref{variationofg} as a function of $a$.  We can see how the sign of $\nu$ determines the increase ($\nu < 0$) or decrease ($\nu > 0$) of the gravitational coupling with the expansion of the Universe, with an overall slow convergence to the present day value of $G_0 \equiv G(H_0)$. 
 This behaviour makes the model compatible with Solar System constraints. In fact, after taking the derivative of Equation \ref{variationofg} one obtains 
\begin{equation}
\frac{\dot{G}}{G} = 2 \nu g(H) (H-\frac{\ddot{a}}{\dot{a}}) < 2 \nu g(H) H,
\end{equation}
where the last inequality holds only in the accelerated expansion epoch where the second term is always positive. 
The most stringent constraints on $\dot{G}/G$ come from Lunar Laser Ranging experiments that give  $|\dot{G}/G| < 0.02 H_0$ \citep{refereetimeg,refereetimeg2}   which, given the range of $\nu$, does not rule out our model on Solar System time-scales.

To fully determine the background evolution for this model we need to rewrite the Friedmann equations in terms of the density parameters, describing the energy densities for matter and the vacuum normalised to the current critical density, $\rho_c^0 = \frac{3H_0^2}{8 \pi G_0}$:
\begin{equation}
    \Omega_i(z) \equiv \frac{\rho_i(z)}{\rho_c^0} ~.
\end{equation}
We choose to express the time-dependence through the redshift $z$ (or equivalently the scale factor $a$) as this will be useful in our analysis. Furthermore, we can define both energy densities normalised to the critical density at an arbitrary redshift, $\rho_c(z) = \frac{3 H^2(z)}{8 \pi G(z)}$, by
\begin{equation}
    \tilde{\Omega}_i(z) \equiv \frac{\rho_i (z)}{\rho_c(z)} = \frac{g(z)}{E^2(z)} \Omega_i(z) ~,
\end{equation}
where $E(z)$ is the Hubble parameter normalised to its current value $H_0$,
\begin{equation}
    E(z) = \frac{H(z)}{H_0} = \sqrt{g(z)} [ \Omega_m(z) + \Omega_\Lambda (z)]^{\frac{1}{2}} ~;
\end{equation}
 its behavior is reported in Figure \ref{fig:handg}. 
It is important to note that in the R-FLRW model the parameters marked with a tilde satisfy the flat space cosmic sum rule at all times, $\tilde{\Omega}_m (z) + \tilde{\Omega}_\Lambda(z) = 1$, while the non-tilde parameters satisfy it only at the present time.

We are now able to write the full system of equations that govern the background expansion in the R-FRLW model:
\begin{align}
    & E^2(z) = g(z)[\Omega_m(z) + \Omega_\Lambda(z)] ~, \label{syshubble}\\
    & (\Omega_m + \Omega_\Lambda) ~\text{d}g + g ~\text{d} \Omega_\Lambda = 0 ~, \label{sysbianchi} \\
    & \Omega_\Lambda(z) = \Omega_\Lambda^0 + \nu [E^2(z) - 1] ~, \label{syslambda} \\
    & \Omega_m(z) = \Omega_m^0 (1+z)^{3(1+w_n)} ~. \label{sysom}
\end{align}
Here, the first equation is the R-FLRW version of the Friedmann equation in the $\Lambda$CDM model, the second is the differential form of the Bianchi equation (Equation~\ref{bianchiconstraint}), the third is just Equation~\ref{secondlambdalaw} rewritten using the density parameter, and the last is a rewrite of the standard equation for $\rho_m$ generalised to include relativistic ($w_m = \frac{1}{3}$) and non-relativistic ($w_m = 0$) matter.

\section{Perturbations}
\label{sec:perturb}
The linear perturbations for the R-FLRW model were thoroughly studied in \cite{Grande_2010}. There, the perturbations of the matter components of the Universe were considered alongside the perturbations for $\rho_\Lambda$ and $G$. This is necessary to grasp the different dynamics present when compared to standard $\Lambda$CDM.   In fact, in both R-FLRW and $\Lambda$CDM, matter is covariantly conserved and the matter density contrast, $\delta_m \equiv \delta\rho_m/\rho_m$, satisfies the following second-order differential equation:
\begin{equation}
\label{scalefactorfirst}
    \delta_m''(a) + \left( \frac{3}{a} + \frac{H'(a)}{H(a)} \right) \delta_m'(a) = \frac{3 \tilde{\Omega}_m(a)}{2 a^2} \left( \delta_m(a) + \frac{\delta G}{G} \right ) ~.
\end{equation}
Here, a prime denotes differentiation with respect to the scale factor. Equation~\ref{scalefactorfirst} reduces to the standard $\Lambda$CDM equation for the scale factor if we assume a vanishing $\delta G$, but to follow the true evolution of the perturbations in a R-FLRW Universe we will assume $\delta G \not= 0$. Then, the perturbations for $\rho_\Lambda$ and $\rho_m$ are related to $\delta G$ through the constraint imposed by the Bianchi identity (Equation~\ref{bianchiconstraint}),
\begin{equation}
\label{perturbationrelations}
    \delta_\Lambda \equiv  \frac{\delta \rho_\Lambda}{\rho_\Lambda} = - \frac{\delta G}{G} ~, ~~~
   \delta_m = - \frac{(\delta G(a))'}{G'(a)} ~.
\end{equation}
  
We can now use these relations to substitute the $\delta G$ terms into Equation~\ref{scalefactorfirst} . After another differentiation we obtain a third-order differential equation for the density contrast that depends only on the cosmological quantities already introduced:
\begin{equation}
\label{scalefactorsecond}
\begin{split}
    &\delta_m'''(a)+\frac{1}{2}\left( 16-9\tilde{\Omega}_m(a) \right) \frac{\delta_m''(a)}{a} + \\
    &\frac{3}{2} \left( 8 - 11 \tilde{\Omega}_m(a) + 3 \tilde{\Omega}_m^2(a) - a \tilde{\Omega}_m'(a) \right) \frac{\delta_m'(a)}{a^2} = 0 ~.
\end{split}
\end{equation}
This equation lends itself to a numerical solution.

\begin{figure}
\includegraphics[width=\columnwidth]{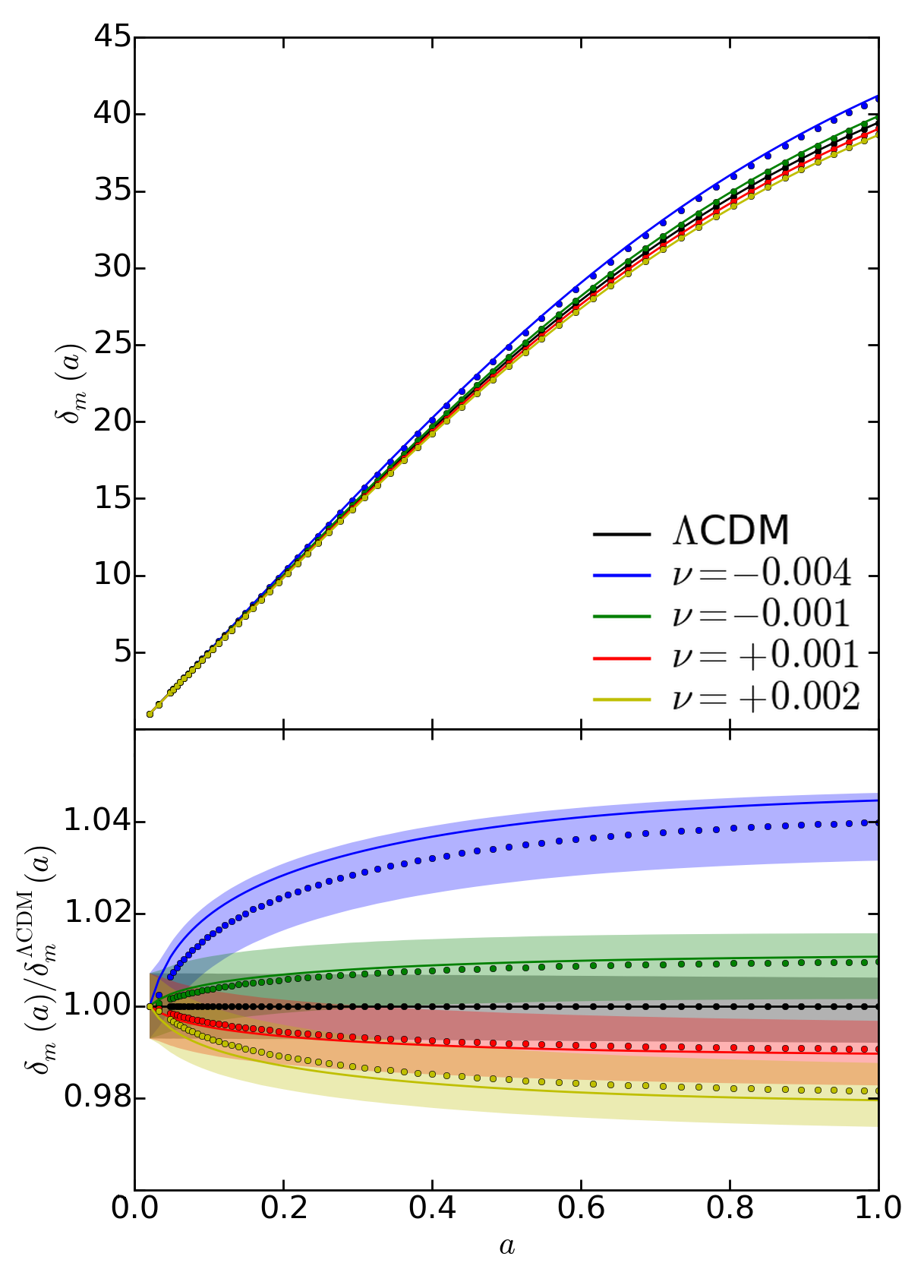}
\caption{In the top panel we show the linear density contrast for different values of the $\nu$ parameter.  The dots represent the linear matter density contrast extracted from the numerical simulations.
In the bottom panel their ratios with  $\Lambda$CDM  are reported, where the shaded areas represent 0.5\% errors around the numerical simulation points.}

\label{fig:growth}
\end{figure}

The behaviour of Equation~\ref{scalefactorsecond} is shown in Figure \ref{fig:growth}. Here we compare the matter density contrast as a function of scale factor for different values of the parameter $\nu$, as well as to the standard $\Lambda$CDM cosmology.
We see that the model predicts an enhancement of growth when $\nu < 0$, due to the strengthening of the gravitational coupling at high redshift that allows the perturbations to overcome the ``repulsion'' associated with expansion, caused by the vacuum energy density. The converse is evident when $\nu > 0$, where the higher value of $\rho_\Lambda$ and the weakening of the gravitational coupling at high redshift hinders the early growth of perturbations.

This behaviour is the main distinguishing feature of the linear analysis of the model. In fact, and as was demonstrated by \cite{Grande_2010}, the shape of the matter power spectrum is the same for the R-FLRW model as that of $\Lambda$CDM.  This is due to the perturbations in $G$ and $\rho_\Lambda$ being negligible at early times.  This allows us to set the power spectrum amplitude for all models at the CMBR redshift $z\sim 1100$ to make them fully compatible with the Planck mission's observations.

\section{Simulations}
\label{sec:simul}

\begin{table}
\begin{center}
\begin{tabular}{lc}
\hhline{==}
Parameter & Value\\
\hline
$\Omega_m$ & 0.3175 \\
$\Omega_\Lambda$ & 0.6825 \\
$\Omega_b$ & 0.0490 \\
$h$ & 0.6711 \\
$n$ & 0.9624 \\
$\sigma_8$ & 0.8344\\
\hhline{==}
\end{tabular}
\caption{Cosmological parameters at redshift $z = 0$ from the Planck mission as reported in \protect\cite{Ade:2015xua}. These have the same definition in both the $\Lambda$CDM and in the R-FLRW models and are used in all our simulations.}
\label{tab:cosmpar}

\end{center}
\end{table}

To explore structure formation in the R-FLRW model we performed a suite of dark matter only N-body simulations.   These simulations follow the evolution of $1024^3$ cold dark matter particles, each of mass $\sim 8\times10^{10} ~M_{\odot}/h$, in a periodic cosmological box of $1024 ~{\rm Mpc}/h$ on a side. From $z=49$ to $z=0$ a total of 62 snapshots were saved.   The suite consists of four simulations that span the natural interval for the $\nu$ parameter as described above, and one control simulation that uses the standard $\Lambda$CDM cosmology. The present day cosmological parameters are the same between the simulations and reflect the latest Planck mission \citep{Ade:2015xua} determination for $\Omega_m$, $\Omega_\Lambda$, $\Omega_b$, $h$, $n$ and $\sigma_8$. These values are reported in Table \ref{tab:cosmpar}.

\begin{table}
\begin{center}
\begin{tabular}{lcc}
\hhline{===}
Simulation & $\sigma_8(z = 0)$ & $\nu$ \\
\hline
H1 & 0.8916 & -0.004 \\
H2 & 0.8479 & -0.001 \\
H3 & 0.8215 & +0.001\\
H4 & 0.8090 & +0.002 \\
\hhline{===}
\end{tabular}
\caption{The different R-FLRW simulations performed. The three columns are the simulation name, the value of $\sigma_8$ at redshift $z=0$ reached by each scenario and the value of the $\nu$ parameter of the R-FLRW model simulated.}
\label{tab:sims}
\end{center}
\end{table}

The simulations were carried out using a modified version of the parallel TreePM N-body code \textsc{gadget-3} \citep{Springel_2005}. This version keeps the original algorithms that evolve the dark matter particles but interpolates the cosmological quantities $H(z)$, $G(z)$, $\Omega_m(z)$, and $\Omega_\Lambda(z)$ using look-up tables. This addition is needed because such quantities now evolve differently in a R-FLRW Universe compared with the standard $\Lambda$CDM that \textsc{gadget-3} usually assumes. The shift to look-up tables also helps make numerical implementation of the model more manageable and avoids an otherwise inevitable performance hit at every time-step.

To populate the look up tables it is necessary to rewrite some of the main equations of the model in a form better suited for numerical integration.  In particular Equation~\ref{syslambda}, when combined with  Equation~\ref{syshubble}, becomes
\begin{equation}
\label{lambdarewrite}
    \Omega_\Lambda(z) = \frac{\Omega_\Lambda^0 + \nu (\Omega_m g(z) - 1)}{1- \nu g(z)} ~,
\end{equation}
which can be rewritten as 
\begin{equation}
\label{lambdarewrite2}
    \Omega_m(z) + \Omega_\Lambda(z) = \frac{\Omega_m(z) + \Omega_\Lambda^0 - \nu}{1- \nu g(z)} ~.
\end{equation}
We then differentiate Equation~\ref{lambdarewrite} to obtain $d\Omega_\Lambda(z)$ and substitute it, along with Equation~\ref{lambdarewrite2}, into Equation~\ref{sysbianchi}. The result is
\begin{equation}
    (\Omega_m(z) + \Omega_\Lambda^0 - \nu) dg + \nu (1 - \nu g) g^2 d\Omega_m(z) = 0 ~,
\end{equation}

which itself can be integrated by quadrature to give $g$ as an implicit function of redshift:
\begin{equation}
    \frac{1}{g(z)} - 1 + \nu \text{ ln } \left [ \frac{1}{g(z)} - \nu \right] = \nu \text{ ln } [\Omega_m(z) + \Omega_\Lambda^0 - \nu] ~.
\end{equation}
It is now straightforward to solve this numerically for $g(z)$ and use the previous equations, along with Equation~\ref{syshubble}, to determine the cosmological quantities needed by the code. We repeat this procedure four times for each of the four values of the $\nu$ parameter (Table~\ref{tab:sims}). Finally, a table was created with the standard $\Lambda$CDM values to allow the use of the same code for all five simulations.

Next, in order for the N-body code to correctly calculate the potential for each of the new models additional changes need to be made. If we consider a perturbation to a spatially flat FLRW metric in the Newtonian gauge,
\begin{equation}
ds^2 =-(1+2\Psi)dt^2+a^2(t)(1-2\Phi)\delta_{ij}dx^i dx^j ~,
\end{equation}
the Einstein equations give $\psi = \phi$, and in the Newtonian limit the perturbation variable $\phi$ plays the role of the gravitational potential. As shown in \cite{Grande_2011} for the R-FLRW model, for deep, sub-Hubble perturbations the potential can be described by
\begin{equation}\label{potentialone}
\phi = - \frac{3}{2} \frac{H^2a^2} {k^2} \left \{ \tilde{\Omega}_m \delta_m + \tilde{\Omega}_\Lambda \delta_\Lambda + \frac{\delta G}{G} \right \} ~,
\end{equation}
 where $\delta_m$ and $\delta_\Lambda$ are the density contrasts defined in Equation \ref{perturbationrelations}.  
This is similar to what must be solved by a standard $\Lambda$CDM N-body algorithm, but with the addition of two extra terms on the end. To include these terms in our code we first rewrite Equation~\ref{potentialone} as
\begin{equation}
\phi = - \frac{3}{2} \frac{H^2a^2} {k^2} \left \{ \tilde{\Omega}_m \delta_m +  (1 - \tilde{\Omega}_\Lambda) \frac{\delta G}{G} \right \} ~,
\end{equation}
where we have used Equation~\ref{perturbationrelations} to get rid of $\delta_m$.   Next we propose the approximation
\begin{equation}\label{eq:ansatz}
\frac{\delta G}{G} = f(a)\frac{\tilde{\Omega}_m}{1-\tilde{\Omega}_\Lambda}\delta_m ~,
\end{equation}
which allows us to rewrite the gravitational potential as
\begin{equation}
\phi = - \frac{3}{2} \frac{H^2a^2} {k^2} (1+ f(a)) \tilde{\Omega}_m \delta_m ~.
\end{equation}
With this, the gravitational potential gains a new temporal dependence through an unknown function $f(a)$. Now the potential can be rewritten as 
\begin{equation}
\label{tildegdef}
\phi = - \frac{4 \pi \tilde{G}(a)}{k^2}\rho_m \delta_m ~,
\end{equation}
which is the standard Newtonian potential with an effective gravitational constant $\tilde{G}(a) = G(a)(1+f(a))$ having an additional temporal dependence through $f(a)$, the effect of which is reported in Figure \ref{fig:handg}.  We stress that, even with this additional contribution to the variation of $G$, the Solar System constrains are still satisfied since $\tilde{G}$ remains slowly-evolving at the present day.\  
With this approximation both the model and perturbations can be easily integrated into any standard N-body algorithm by adjusting $G$ at every timestep.

To find a functional form for $f(a)$ we first solve numerically Equation~\ref{scalefactorsecond} for $\delta_m$ and then plug this solution into the second of Equation \ref{perturbationrelations}, which is again numerically solved to give $\delta G(a)$. This can then be used in Equation \ref{eq:ansatz}, which defines our proposed approximation, to obtain a numerical solution for $f(a)$.\
This derivation step was repeated for all four values of the $\nu$ parameter, and the result was included in the pre-computed tables fed to the N-body code.

We then put this approximation to the test in two ways. First, we plug it into Equation \ref{scalefactorfirst} and solve for $\delta_m$ to check that it does not differ by more than 1\% from the $\delta_m$ previously obtained when solving the exact equation, Equation~\ref{scalefactorsecond}.  The second test is performed after the simulations are completed, by comparing the growth of a large scale mode in the measured power spectrum of every snapshot in our simulations against the solution for $\delta_m$ obtained from the exact equation. The result of this second test is shown in Figure \ref{fig:growth}, where we see how the simulations (symbols) provide a better than 0.5\% (shaded areas) agreement for the linear growth of structure when compared with the exact solution (solid lines).  
\subsection{Initial Conditions}
Finally, the last part of the simulation pipeline that required careful consideration given the change in cosmology was the generation of the initial conditions. These were obtained by perturbing a glass particle distribution according to the 2LPT prescription described by \cite{Crocce_2006} using the 2LPTic code. This code needed some minor tweaks to take into account the modified evolution of the cosmological quantities, but used only two values for each of them, one at the starting redshift, chosen to be $z=49$, and one at the present time. These values were calculated using the look-up tables described before.

To generate the initial conditions we need to draw the phases from a random distribution given a predefined shape and amplitude of the linear power spectrum. We initialize the distribution using the same random seed across the different scenarios and use the same power spectrum shape obtained with the powerful CAMB code described in \cite{Lewis:2002ah}. This code calculates an accurate $\Lambda$CDM linear power spectrum shape at any redshift given the cosmological parameters at redshift $z=0$. We can use the same calculation for our R-FLRW simulations since, as discussed above, this model retains the same linear power spectrum shape as that of the $\Lambda$CDM model.   The amplitude of the power spectrum is then set according to the $\sigma_8$ value given in Table \ref{tab:cosmpar}, scaled back to the epoch of recombination ($z = 1100$) using the standard $\Lambda$CDM formula for the growth factor, and then scaled forward to the initial redshift of our simulations, $z=49$, using the numerical solution of equation Equation~\ref{scalefactorsecond}.   This choice is equivalent to normalising the power spectrum of every realisation to the same $\sigma_8$ at the epoch of recombination and results in a different $\sigma_8$ at redshift $z=0$ for every scenario, as reported in Table \ref{tab:sims} . This, as we discussed in Section \ref{sec:perturb}, makes our models compatible with the Planck mission's observations and allows us to focus our analysis on the resulting differences measurable at low redshift.

\subsection{Halo Finding}
Our version of the \textsc{gadget-3} code performs two levels of halo identification for every snapshot saved. These routines inherit the modifications previously discussed, but aside from this were otherwise unchanged. First, the Friends-Of-Friends (FOF) algorithm \citep{Davis_1985} used by \textsc{gadget-3} identifies halos based on a nearest neighbour search, with a linking length $b=0.2$ of the mean inter-particle separation. The mean density of such halos approximately correspond to the overdensity of virialised structures expected from the spherical collapse model. Second, substructures are then traced using the \textsc{subfind} \citep{Springel_2005} algorithm that groups gravitationally self-bound particles around local density maxima so that every FOF group contains at least one sub-halo.
\subsection{Merger Tree Construction}
Once each simulation had run and all (sub)structures identified and measured, halos were then linked across the output snapshots using the \textsc{l-halotree} code to construct the merger tree of each $z=0$ object. Such trees describe the evolution of mass and other halo properties with time.

\section{Results and Discussion}
\label{sec:results}
\subsection{The non-linear Power Spectrum}
\label{sec:pk}
\begin{figure*}
\includegraphics[width=\textwidth]{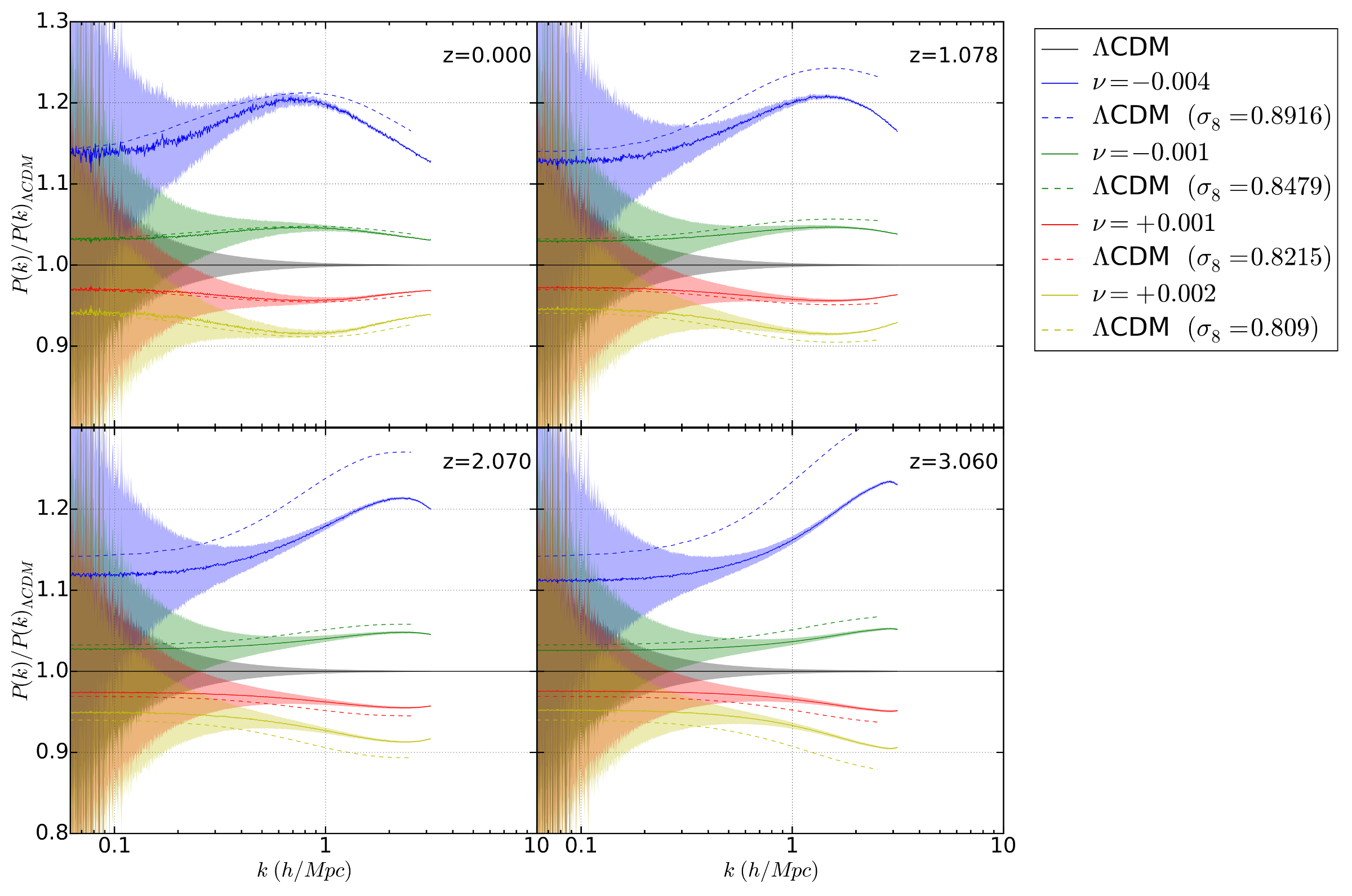}

\caption{The ratio of the non-linear power spectrum for each R-FLRW simulation to the non-linear power spectrum of the $\Lambda$CDM simulation. The colour coding shows the different values of the $\nu$ parameter used for the R-FLRW model simulations, as defined in the legend. The dashed lines represent the same ratio calculated for a $\Lambda$CDM model with the same $\sigma_8$ as the corresponding simulation, as described in Section~\ref{sec:pk}. }
\label{fig:ps}
\end{figure*}

We begin our analysis of structure formation in the non-linear regime by computing the power spectra of all our simulations at various redshifts. This enables us to quantify the differences in the density field between our four R-FLRW simulations and the $\Lambda$CDM simulation. To compute the density power spectrum, $P(k)$, we used the routines built into the N-body code, since these have access to the density field already computed for the calculation of the gravitational force. This method provides a robust determination of the power spectrum on non-linear scales up to the Nyquist frequency, $k \sim 3.2$ $h$/Mpc. These realisations of the non-linear power spectrum represent a first probe of the impact on the clustering of matter from the variation of the gravitational and cosmological ``constants'' in this scenario, and extends our previous analytical treatment which was limited to the the overall growth on linear scales only.

Since the R-FLRW model does not exhibit any significant departures from the $\Lambda$CDM model in the early Universe, when setting the initial conditions we assumed the same density perturbation normalisation for all simulations at the redshift of the CMBR, $z \approx 1100$. The consequence of this choice is that, because the subsequent density evolution in each simulation will be different, by redshift zero each will end at a different value of $\sigma_8$. Therefore, from linear theory alone we expect the linear part of the power spectrum in the R-FLRW simulations to have the same shape as the $\Lambda$CDM simulation but exhibit a different normalisation across all redshifts. To highlight these differences, in Figure~\ref{fig:ps} we plot the ratio of the power spectrum of each R-FLRW simulation to the $\Lambda$CDM simulation, $P(k)/P(k)_{\Lambda CDM}$. The shaded areas represent the shot noise due to the finite number of particles used, and the power spectra have been cut where the shot noise reaches 10\%.

In Figure~\ref{fig:ps} we see a clear, constant difference on the largest scales, $k \lesssim 0.2$, for all R-FLRW simulations. The two $\nu < 0$ scenarios simulated, which end with a larger $\sigma_8$, show a power spectrum amplitude that is 13\% and 3\% higher than the $\Lambda$CDM simulation. In contrast, the $\nu > 0$ simulations end with a lower $\sigma_8$, and this results in a power spectrum amplitude 3\% and 5\% lower than that of the $\Lambda$CDM simulation.

On smaller scales all R-FLRW models exhibit an enhancement of the large-scale behaviours, emphasised by an apparent peak at around $k \sim 1$. For the $\nu = -0.004$ case this reaches a maximum 20\% difference, and for $\nu = -0.001$ a 5\% difference. The same happens for the $\nu > 0$ cases but in the opposite direction, where the amplitude becomes 5\% and 8\% lower that the $\Lambda$CDM case at the peak. These features become even more prominent at higher redshift where the amplitude difference on the smallest scales for the most extreme scenarios simulated reaches the 22\% and 20\% at redshift $z \sim 3$, respectively, while the peak amplitudes in the intermediate scenarios reach a maximum difference of 5\%.

Beyond the normalisation of the power spectrum, to understand how the new dynamics within each model affect the formation of structure, we calculate the non-linear power spectrum of a $\Lambda$CDM cosmology with the same value of $\sigma_8$ at redshift zero that each R-FLRW model ends up at. This can be accomplished without a new simulation thanks to the HALOFIT \citep{halofitpap} procedure implemented in the CAMB \citep{CAMBpap} code. The result is then compared to the vanilla $\Lambda$CDM cosmology from before by again taking the ratio of each result. This allows us to explore the differences at each $k$ between pure $\Lambda$CDM with the final R-FLRW $\sigma_8$, and the R-FLRW simulations themselves that end at that $\sigma_8$. These curves are over-plotted in Figure~\ref{fig:ps} by the series of dashed lines in each panel, as labelled.

Considering the dashed curves in Figure~\ref{fig:ps} and comparing them to the solid curves, we see how they share many of the same features. At redshift zero the power spectra for all simulations except $\nu = -0.004$ are almost completely degenerate with each comparable $\sigma_8$ $\Lambda$CDM result. The $\nu = -0.004$ simulation, instead, shows an amplitude that is a few percent lower. To break these degeneracies it is necessary to examine the high redshift power spectra, and here we find a clear departure of $\Lambda$CDM from the R-FLRW simulations. In particular, the amplitude differences increase with increasing redshift. For example, a $\Lambda$CDM model with the same $z = 0$ $\sigma_8$ as a $\nu < 0$ R-FLRW model will have similar clustering at redshift zero but will be more clustered in the past. For $\nu = -0.004$, in fact, we see a $\sim$10\% lower power spectrum amplitude by $z = 3$. Conversely, a $\nu > 0$ R-FLRW universe is always less clustered at $z = 3$ than the corresponding $\Lambda$CDM model, but with departures of only a few percent. Both the extreme positive and negative $\nu$ cases show differences that lie abundantly within the precision range sought after by future cosmological experiments, such as Euclid \citep{EuclidPap}. For the intermediate cases, while we find small differences, they might prove to be very difficult to distinguish. Overall the R-FLRW model power spectra exhibit interesting departures from $\Lambda$CDM that may provide discriminatory power.

\subsection{Halos} \label{sec:halos}
\begin{figure*}
\includegraphics[width=\textwidth]{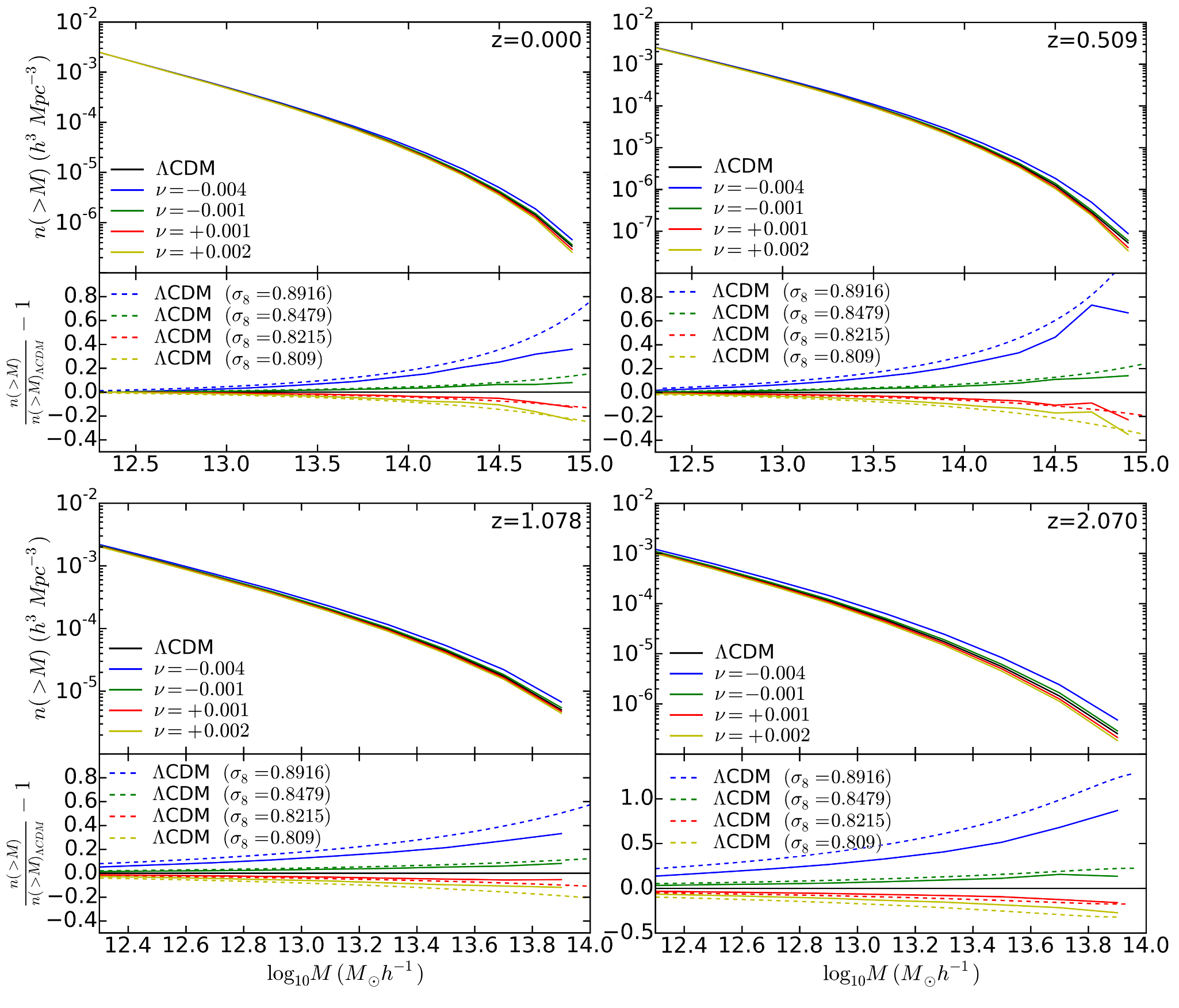}

\caption{The halo mass function for our five simulations at four redshifts, from $z=0$ to $z=2$. The colour coding follows the convention of the previous figures, as marked. In the lower section of each panel we plot the difference relative to the $\Lambda$CDM halo mass function. The dashed lines represent the same ratio calculated for a $\Lambda$CDM model with the same $\sigma_8$ as the corresponding R-FLRW simulation. }
\label{fig:hmf}
\end{figure*}

Next we study the impact of the modified R-FLRW dynamics on the statistics of the halo population. In \cite{Grande_2011} it was shown, using Press-Schecter theory, how a R-FLRW scenario with $\nu = -0.004$ produces an increase in the expected number of high mass halos at high redshift. We expand the predictions made in that work across a larger halo mass range, between $10^{12}$ to $10^{15} M_{\odot}/h$, and a larger redshift range, up to $z \sim 2$. In Figure~\ref{fig:hmf} we plot the cumulative halo mass function, $N(> M)$, for each of our five simulations. To remain above the simulation mass resolution limit we only include halos with masses larger than $10^{12} M_{\odot}/h$; such halos are resolved with at least 20 dark matter particles by the halo finder.

Considering each solid curve in Figure~\ref{fig:hmf}, we find a similar number density of low mass halos and a small divergence at high masses for all the R-FLRW models plotted. In the $\nu = -0.004$ case at $z = 0$, the difference of this mass function to the $\Lambda$CDM simulation reaches a value of about 40\% for halos larger than $10^{14.5} M_{\odot}/h$. This becomes more significant at intermediate redshifts, with the abundance of massive halos at $z = 0.5$ more than 50\% larger than the Planck $\Lambda$CDM case.

Moving to even higher redshifts, such massive cluster-sized halos have not yet formed in the simulations, so the excess abundance shifts to lower masses. In the $\nu = -0.004$ R-FLRW model it reaches a 40\% excess for halos more massive than $10^{13.7} M_{\odot}/h$ at $z = 1$, and a 70\% excess for halos more massive than $10^{13.8} M_{\odot}/h$ at $z = 2$. The intermediate R-FLRW models, while showing similar trends, never reach excesses larger than 10\% across all masses, and the same can be said for the other extreme case, $\nu = 0.002$, which stays well below a 15\% excess, even at the highest redshifts considered.

As we discussed in the previous section, it is important to characterise how much of these differences is due to the different linear growth so that we can single out the contribution from the new dynamics. To do so we use a \cite{tinkerfit} fitting function implemented in the HMFcalc \citep{hmfcalcpap} tool. This fitting function gives us a cumulative number density of halos for a $\Lambda$CDM model with the parameters listed in Table \ref{tab:cosmpar}. As before, we only change the value of $\sigma_8$ at redshift $z = 0$ so as to match the value reached by each of the R-FLRW simulations. Again we take the ratio of each fitting function to that from the $\Lambda$CDM simulation. These new ratios are plotted in the lower panels of Figure~\ref{fig:hmf} with dashed lines.

At $z = 0$ we see very similar behaviour (dashed lines) to the R-FLRW curves (solid lines), where the differences are typically less than a few percent for all simulations except when $\nu = -0.004$. There we find a larger departure from its $\Lambda$CDM equivalent, approaching a $\sim 10$\% lower number density at the highest masses. Moving to higher redshifts, these differences in the ratios remain minor, although by $z = 2$ more significant deviations across all masses are found again for $\nu = -0.004$.

Future observational probes will measure the halo mass function at high redshift with great precision. Coupling these observations with a robust determination of $\sigma_8$ should allow astronomers to differentiate the different R-FLRW models, which are otherwise degenerate at $z = 0$.

\subsection{Subhalos}
\label{sec:subhalos}

\begin{figure*}
\includegraphics[width=\textwidth]{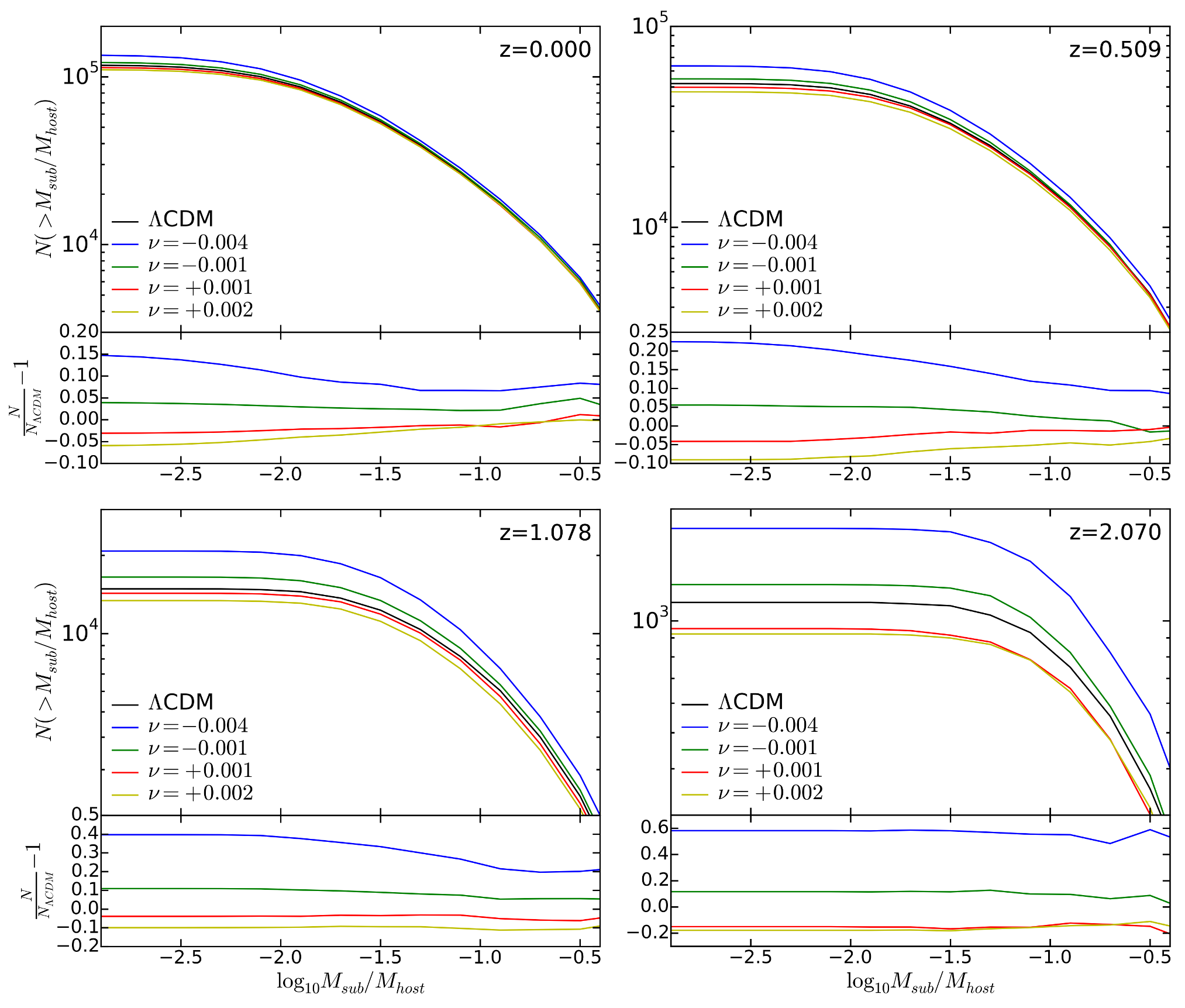}

\caption{The SubHalo Mass Function the different simulations at different redshifts. The colour coding follows the convention of the previous figures, as marked. In the lower section of each panel we plot the difference relative to the $\Lambda$CDM subhalo mass function. }
\label{fig:subhmf}
\end{figure*}

To examine the effects of the R-FLRW model on the internal sub-structure of dark matter halos we calculate the cumulative subhalo mass function. This function counts the number of subhalos within a halo that share a given fraction of the total host halo mass. We include only hosts that contain at least 5 subhalos and select only well resolved hosts that are composed of at least 100 simulation particles. In Figure \ref{fig:subhmf} the results for all five R-FLRW models are presented, at the same four redshifts as before. In the lower part of each panel we plot the ratio of each model to the $\Lambda$CDM model, highlighting their relative differences.

At $z = 0$ the models are remarkably close to each other and the differences have a very shallow dependence on the fraction of mass shared with the host. The maximum difference is again reached in the $\nu = -0.004$ scenario, with up to 15\% more substructures for $M_{sub}/M_{host} > 10^{-2.5}$ than the $\Lambda$CDM model. This difference drops to 10\% as we move to larger fractions. For the other R-FLRW models the differences are constantly below 5\%. Overall, the $\nu < 0$ case shows an increase in the number of subhalos at a given mass fraction, particularly at small mass ratios, while $\nu > 0$ shows the opposite trend.

At higher redshifts the deviations for the $\nu = -0.004$ scenario from Planck $\Lambda$CDM become stronger, reaching up to 20\% at $z = 0.5$, 40\% at $z=1$, and 60\% at $z=2$. The other scenarios reach a maximum difference of 15\% by redshift 2. Furthermore, the subhalo fraction dependence becomes shallower at higher redshift for all models, and is almost constant by $z=2$.

This behaviour can have important consequences for galaxy formation, leading, in the case of $\nu = -0.004$, to an excess in the number of satellite galaxies around a given galaxy when compared with a $\Lambda$CDM cosmology. This excess persists all the way to the present day, in particular for low mass fractions, hinting at an expected larger number of local dwarf galaxies. This could perhaps exacerbate the missing satellite problem \citep{mooresatellite, klypinsatellite}. In contrast, at high redshift the opposite happens in the $\nu > 0$ cases. However, while showing close to 20\% less subhalos at high redshift, these models end up with a subhalo abundance difference smaller than 5\% when compared to $\Lambda$CDM by redshift $z=0$. This difference might be observationally difficult to measure, even in future surveys. The consequences of the various R-FLRW models on galaxy formation will be studied quantitatively in a future publication.

\subsection{Mass History}
\label{sec:mah}

\begin{figure*}
\includegraphics[width=\textwidth]{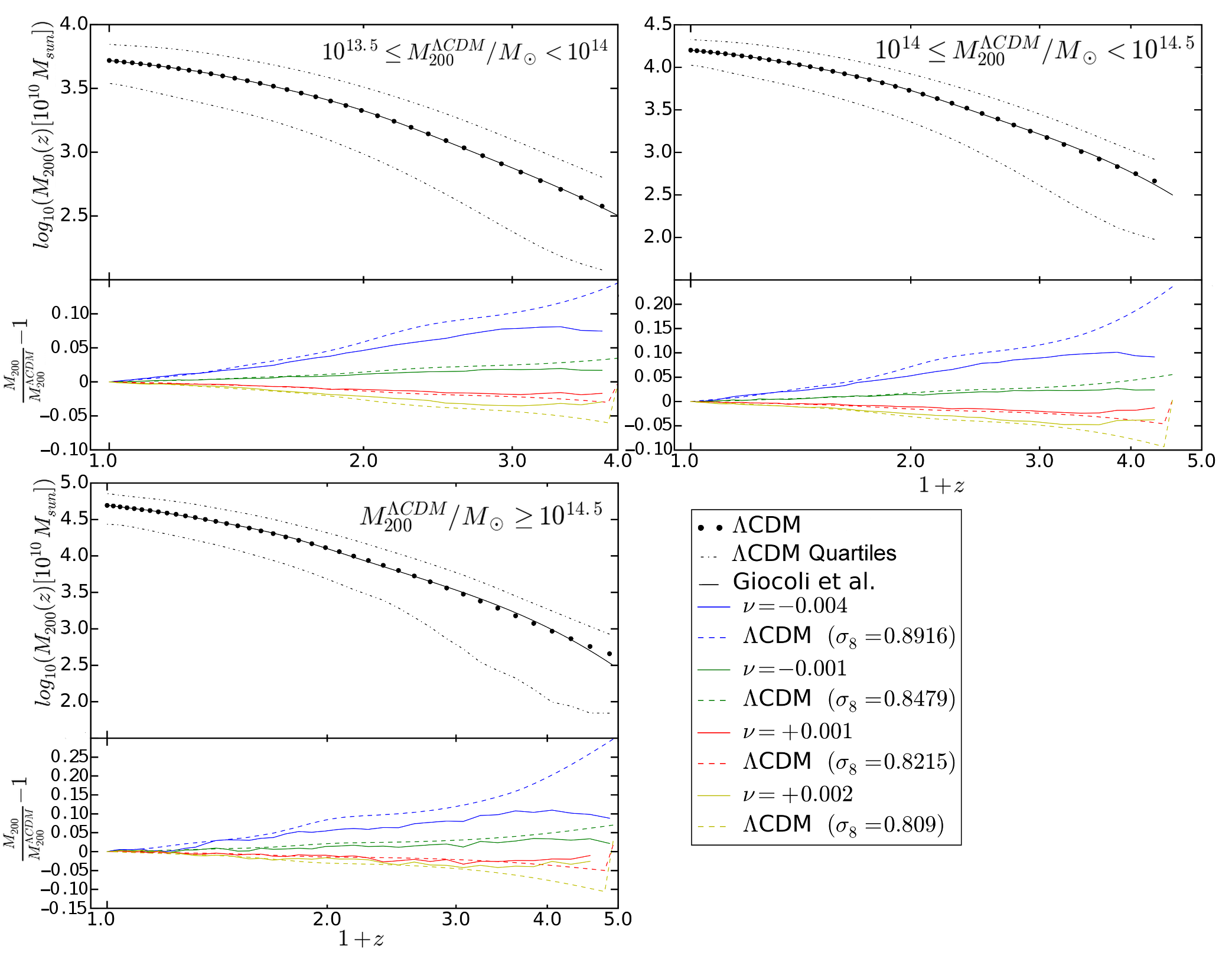}

\caption{The  Mass Accretion History for the different simulations is represented as the median halo mass as a function of redshift. The dashed lines in the upper section of each panel represent the first and last quartiles of the distribution of mass histories. In the lower section of each panel we plot the difference in the mass accretion history for every R-FLRW simulation relative to the $\Lambda$CDM one. The colour coding follows the convention of the previous figures, as marked. }
\label{fig:mah}
\end{figure*}

Our simulation post-processing includes both halo finding and merger tree construction. The merger tree code links every halo with its progenitors in the previous snapshot. This is accomplished by following all the individual particles that are bound to a halo in one snapshot and checking to which halo they belonged in the previous. In this way we can follow the evolution of the properties of a structure across time and identify accretion events, such as mergers.

In particular, in this section we are interested in how halos grow and how the new R-FLRW dynamics affect this. We can build a halo's mass accretion history by following the progenitor links, starting at $z=0$, back in time to where the halo was first identified in the simulation. This history, denoted $M(z)$, was calculated for every halo. By taking the median of all $M(z)$ we obtain an average picture of how accretion proceeds in each R-FLRW model.

We first divide the halos at $z=0$ into three mass bins and calculate their median growth history. The three panels in Figure \ref{fig:mah} report the results for the $\Lambda$CDM simulation in each of these mass ranges, as labelled, including the first and last quartiles of the distribution. In the lower part of each panel we plot the ratio between the $\Lambda$CDM and R-FLRW simulations. We can see how the halos in the $\nu < 0$ scenarios tend to be more massive earlier in time, while the opposite is true for $\nu > 0$, with smaller halos at higher redshift. This behaviour is evident for all three mass ranges, showing around 5\% or less deviations. It is important to note that these features are much smaller than the intrinsic scatter in the median $M(z)$ relation.

In the upper part of each panel in Figure \ref{fig:mah} we also plot the model of \cite{Giocoli01102013}, developed to reproduce the halo growth in term of $M_{200}$ in the $\Lambda$CDM model. The model was originally tuned to the \cite{baldiCodecs} simulations so for use here we adjusted its parameters to reflect our updated Planck cosmology. The agreement is remarkably good (circles to solid black line), so we use this model to compare each R-FLRW simulation with a $\Lambda$CDM scenario having equivalent $\sigma_8$, as done in the previous sections. To do so we rescale the \cite{Giocoli01102013} model to the $\sigma_8$ at redshift $z=0$ reached by each R-FLRW simulation. Each result is reported by the dashed lines in the lower part of each panel of Figure \ref{fig:mah}. Again we can see how the trends are reversed; comparing a $\Lambda$CDM Universe having the same $\sigma_8$ as each R-FLRW model, for $\nu < 0$ ($\nu > 0$) the $\Lambda$CDM halos will be, on average, more (less) massive at high redshift. These differences are consistent across all the mass ranges explored but are still small when compared to the intrinsic scatter of the global mass accretion history.

\subsection{Matching}
\label{sec:matching}

\begin{figure*}
\includegraphics[scale=0.35]{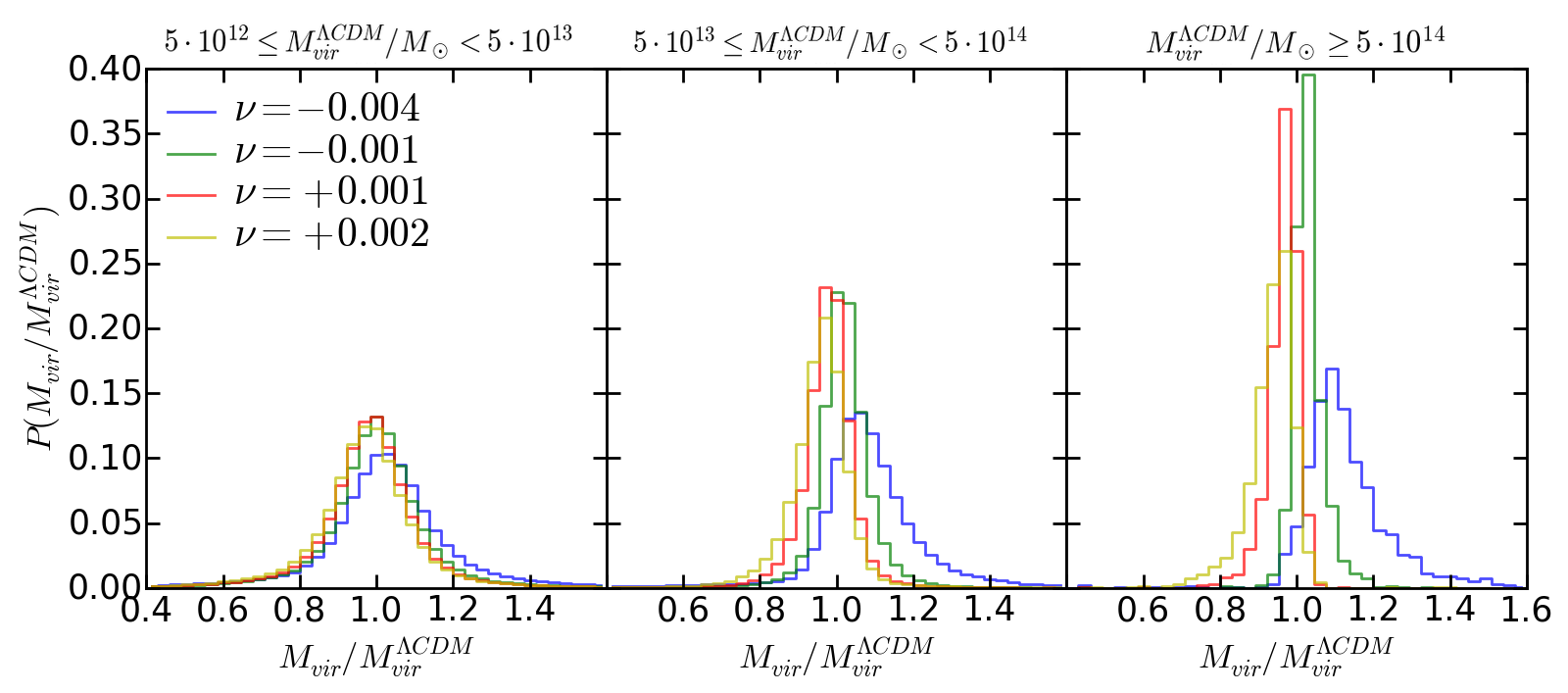}

\caption{The mass ratio for matched halos   at redshift $z=0$  . The colour coding follows the convention of the previous figures. as marked}
\label{fig:matchmass}
\end{figure*}

Because the initial conditions of our simulations all share the same random phases and same power spectrum normalisation at the redshift of the CMBR ($z \simeq 1100$), we can track the movement of the same particles across all the five R-FLRW scenarios and compare them to the $\Lambda$CDM case. In essence, every halo has a ``partner'' halo in each simulation that can be measured and followed. To this end, we modified the merger tree code so that instead of searching for a progenitor across two snapshots of the same simulation it looks for progenitors across the same snapshot in two different simulations. Then, instead of considering many progenitors we restrict ourselves to finding the main progenitor and call this halo a ``partner''. This is, by definition, the structure that shares the largest number of particles with the one of interest in $\Lambda$CDM simulation. Our procedure also ensures that every structure in the $\Lambda$CDM simulation will have only one counterpart in every R-FLRW simulation. Once we obtain a catalog of these matching structures across all our simulations we can compare the properties of an individual $\Lambda$CDM halo to its partner in each R-FLRW simulation.

In Figure~\ref{fig:matchmass} we show the distribution ratio of matched halo virial mass in a R-FLRW cosmology to that in the $\Lambda$CDM cosmology, $M/M_{\Lambda CDM}$, for each different R-FLRW simulation (different lines). We calculate separate distributions for three mass bins shown in the three panels, as labeled. In each mass bin we find that the distributions are close to Gaussian, especially for the lowest mass bin.   In this bin, each R-FLRW cosmology peaks near a ratio of 1. Furthermore, the distribution width for low mass halos is remarkably similar for each comparison, but this is driven by fact that such halos are near the simulation resolution limit and the small number of particles needed to resolve them makes the possible variations in their  mass very small.  

As we move to the intermediate mass bin the peaks of each curve become sharper, in particular for the less extreme R-FLRW cosmologies. The matched mass distributions for these three have almost perfect overlap, with a narrower spread than the lower mass bin but again centred on ratios of unity. The most extreme simulation, $\nu = -0.004$, on the other hand, shows a distribution with a shallower peak that is slightly skewed towards higher ratios by about 5-10\%. In the highest mass bin the distributions become even narrower and the peaks become even sharper. We find that the $\nu < 0$ ($\nu > 0$) results shift further away from unity to higher (lower) ratios, typically by about 10\%.

The mass ratio distributions in Figure~\ref{fig:matchmass} indicate that higher mass halos have a higher probability of ending up more massive in a $\nu = -0.004$ R-FLRW cosmology when compared to the $\Lambda$CDM case. On the other hand, the intermediate R-FLRW scenarios are morel likely to result in halos of very similar mass, or even a little lower. These results support what was found in Section~\ref{sec:halos} when we examined the halo mass functions of each R-FLRW Universe. There, taking $\nu < 0$ ($\nu > 0$) resulted in a higher (lower) number density of high mass halos when compared to $\Lambda$CDM.

 Using the same procedure we also looked for variations in the concentration of halos but found no significant differences in the distribution between $\Lambda$CDM and the simulated R-FLRW scenarios. 

Our matching procedure also provides a window into the differing formation histories that halos can have in a modified gravity scenario. Using the mass accretion history discussed in the previous section we calculate the formation time of each halo found at $z=0$ and compare them to the formation time found in the $\Lambda$CDM simulation for the same halo. For our work, we define formation time as the time when a structure has obtained a set fraction of its total final mass \citep{Giocoli01102013}: 90\%, 50\%, 10\% and 4\%, denoted as $t_{0.9}$, $t_{0.5}$, $t_{0.1}$, $t_{0.04}$ respectively. We then calculate the difference between the formation time of each $\Lambda$CDM halo and its corresponding halo in each R-FLRW simulation. 

When comparing formation times, although trends could be seen in each distribution we found that the mean differences for matched halos were always small, of order 1-10 Myr. Given the time resolution of our simulations are typically an order-of-magnitude larger than this we mark these deviations as interesting but too small to claim as significant with the current number of snapshots. We highlight though that formation time remains an interesting probe for future theoretical and observational work, assuming a sufficient level of accuracy can be reached.

\section{Conclusions}
\label{sec:conclusions}

In this paper we introduced the ``Running FLRW'' cosmological model and described the numerical set-up devised to perform a suite of cosmological N-body simulations aimed at investigating structure formation in this alternative cosmological scenario. 

The R-FLRW model was first discussed in \cite{Grande_2011}, where they show how the running of the \emph{cosmological constant} in a $\Lambda$CDM model can be interpreted as having the same effective behaviour as seen in vacuum models from quantum field theory. In the present work, we showed how an additional requirement that matter be covariantly conserved results in the running of the \emph{gravitational constant} as well. We described the background expansion and the evolution of linear perturbations with particular attention to the numerical implementation. This consisted of numerical solutions to the set of equations for the common cosmological quantities, like the Hubble parameter $H(a)$ and the density contrast $\delta_m(a)$, taking into account a variable cosmological ``constant'' $\Lambda(H)$ and gravitational ``constant'' $G(H)$. The magnitude of this time dependence is conveyed by the $\nu$ parameter introduced Section \ref{sec:model}, and the choice of the parameter values for our simulations was made to span the 1-$\sigma$ interval already constrained by \cite{Grande_2011} using CMBR, BAO and supernovae observations. All the cosmological functions were then derived beforehand for every value of the $\nu$ parameter while keeping the other cosmological parameters from Table \ref{tab:cosmpar} fixed. These were then incorporated in the N-body code calculations through the interpolation of tabulated values.

All functional modifications included a correction derived from perturbation theory to take into account the perturbations in $G$ and $\Lambda$ that are not otherwise present in the $\Lambda$CDM model. To include these perturbations, in Section \ref{sec:perturb}, we proposed an ansatz, given by Equation~\ref{eq:ansatz}, and showed how it provides a good approximation to the exact analytical solution.

The final step in the set-up of the simulation suite involved the choice of the normalisation of the power spectrum for the initial conditions. Following \cite{baldiCodecs}, we chose to identically normalize every simulation at the redshift of the last scattering surface, $z \simeq 1100$. This is possible thanks to the work of \cite{Grande_2010} who demonstrated that the running of $\Lambda$ and $G$ does not alter the transfer functions of the $\Lambda$CDM model. Our choice ensured that the shape of the power spectrum was the same across all our simulated scenarios.

After completing the R-FLRW simulation suite and a reference simulation with a standard $\Lambda$CDM cosmology, we examined the features of each, with close attention to the impact of the R-FLRW scenario on structure formation in the non-linear regime. We began our analysis with a comparison of the non-linear power spectrum. Each R-FLRW simulation, being distinguished by a different evolution of the linear density contrast, ends up with a different value of $\sigma_8$ at redshift 0. As discussed in Section \ref{sec:pk} our results recover the expected normalisation difference in the linear part of the power spectrum. However the non-linear part shows deviations from $\Lambda$CDM as large as 20\% at redshift $z=0$, growing to 30\% by redshift $z=3$.

To investigate the degeneracy of $\nu$ with $\sigma_8$ we compared each R-FLRW realisation with a $\Lambda$CDM non-linear power spectrum having the same $\sigma_8$, calculated using the HALOFIT routine. We see a high level of degeneracy at redshift 0 for all the realisations, with the more extreme models exhibit some interesting departures at very small scales. To break this degeneracy we find it necessary to exploit the redshift evolution of these differences. Here we use the fact that clustering in the R-FLRW model proceeds at a different pace than in a $\Lambda$CDM cosmology with the same $\sigma_8$, being slower for $\nu < 0$ and faster for $\nu > 0$. This might be a `smoking gun' of the model, exploitable in future large-scale structure surveys like Euclid \citep{EuclidPap}.

For the second part of the paper we focused on the halo populations in each R-FLRW scenario. This allowed us to extend the Press-Schechter theory analysis performed by \cite{Grande_2011} in both parameter and redshift space. In Section \ref{sec:halos} we discussed how our simulations revealed a significant difference in the number of halos at high halo mass, in particular for the most extreme values of $\nu$. These differences grow with redshift and exceeded $\Lambda$CDM by as much as 50\% for $\nu = -0.004$. Smaller differences were found for smaller halo masses, which were again most prominent at higher redshift.

As before, we quantified how much of this effect is due to the differing $\sigma_8$ at redshift $z=0$ for each R-FLRW simulation. This was accomplished with the aid of the HMFcalc tool that uses a Tinker fitting function to analytically determine the halo mass function. We found a strong degeneracy for all  $\nu$ values explored except $\nu = -0.004$. For this more extreme scenario the excess in the number density of halos was $\sim$ 30\% at redshift 2 when compared with a $\Lambda$CDM cosmology having the same $\sigma_8$. These kind of deviations will be highly constrained by future cluster surveys like eROSITA \citep{eROSITA}.

However the most striking difference found in our work was in the subhalo population, where we see an almost constant offset in the number of subhalos as a function of mass fraction. This offset can grow as large as 60\% for the $\nu = -0.004$ scenario but is also significant for the intermediate values. In Section \ref{sec:subhalos} we discussed how such behaviour may have important implications for the satellite galaxy population of halos of all sizes and will be further explored in a future work focused on galaxy formation in the R-FLRW scenarios.

In the final part of the paper we turned our attention to the particle data available from our simulations. This allowed us to link halos across snapshots to build halo merger trees, which can then be used to investigate the mass accretion history of individual objects or clusters. In Section \ref{sec:mah} we examine the distribution of mass histories as a function of halo mass, and while there was very little variation across the simulations, the mean growth histories of high mass halos did present differences of up to the 20\% in the most extreme R-FLRW scenario. 
Our simulations were also compared with the $\Lambda$CDM mass accretion history fitting function of \cite{Giocoli01102013} to check for degeneracies with $\sigma_8$. Again, we find a high degree of degeneracy at low redshift that decreases only at very high redshifts. 

Finally, given that all simulations were run with identical initial conditions, mass and time resolution etc, we undertook a halo-to-halo comparison test between the different cosmological models. In Section \ref{sec:matching} we describe how we modified our merger tree construction procedure to find the same halo in each simulation of our suite. In this way we were able to localise any change in halo properties to the different cosmological scenario alone. We focused on comparing both the mass growth through the halo mass ratio, and the differences in the formation times, to the $\Lambda$CDM simulation. The differences found were statistically small across various properties but showed wide deviations on a halo-to-halo basis, and with an overall agreement to our previous findings.
\\
\\
The aim of this work was to contribute to the current computational efforts being undertaken by the cosmology community looking to understand what dark energy is, not just how it manifests in our Universe. To this end we ran a novel suite of simulations exploring the non-linear regime in the R-FLRW cosmological model, extending the current literature and providing tests against recent cosmological observations and for future surveys. Our simulations use the latest determination of the cosmological parameters, and in their analysis we exposed a number of degeneracies and discussed the methods and observations needed to break them. The results presented here, in combination with future survey data, will help to further test and possibly falsify the R-FLRW scenario.

\section*{Acknowledgments}
We would like to thank Volker Springel and Klaus Dolag for sharing the \textsc{gadget-3}  and \textsc{lhalotrees}  codes. We also thank Steven Murray for making the HMFcalc tool publicly available. We thank the reviewer for their thorough review and highly appreciate the comments and
suggestions, which significantly contributed to improving the quality of the paper.

\bibliography{bibliography/converted_to_latex.bib}

\end{document}